\documentclass[aps,prl,twocolumn,superscriptaddress,showpacs]{revtex4}

\usepackage[dvips]{graphicx}
\usepackage{latexsym}

\begin{document}

\title{Ubiquitous V-shape density of states in mixed state
of clean limit type II superconductors; STM experiment and theory
}

\author{
N. Nakai
}
\email[]{nakai@mp.okayama-u.ac.jp}
\affiliation{
Department of Physics, 
Okayama University, Okayama 700-8530, Japan
}
\author{
P. Miranovi\'{c}
}
\affiliation{
Department of Physics, 
University of Montenegro, Podgorica 81000, Serbia and Montenegro
}

\author{
M. Ichioka
}
\affiliation{
Department of Physics, 
Okayama University, Okayama 700-8530, Japan
}
\author{
H. F. Hess
}
\affiliation{
NuQuest Research LLC,
La Jolla, CA 92037-5801, USA
}
\author{
K. Uchiyama
}
\affiliation{
Department of Physics, 
Tokyo Institute of Technology, Tokyo 152-8551, Japan
}
\author{
H. Nishimori
}
\affiliation{
Department of Physics, 
Tokyo Institute of Technology, Tokyo 152-8551, Japan
}
\author{
S. Kaneko
}
\affiliation{
Department of Physics, 
Tokyo Institute of Technology, Tokyo 152-8551, Japan
}
\author{
N. Nishida
}
\affiliation{
Department of Physics, 
Tokyo Institute of Technology, Tokyo 152-8551, Japan
}

\author{
K. Machida
}
\affiliation{
Department of Physics, 
Okayama University, Okayama 700-8530, Japan
}
\date{\today}

\begin{abstract}
It is demonstrated theoretically and experimentally 
that the low energy density of states $N(E)$ 
is described by a singular V-shape form $N(E)=N_0(H)+\alpha |E|+O(E^2)$
for all clean superconductors in a vortex state, 
irrespective of the underlying gap structure. 
The linear term $\alpha |E|$ which has been not recognized so far
is obtained by exactly evaluating the vortex contribution.
Based on microscopic Eilenberger theory $N(E)$ is evaluated for isotropic gap,
line and point node gaps to yield a V-shape $N(E)$. STS-STM experiments on 
NbSe$_2$ and YNi$_2$B$_2$C give direct evidence for this. 
We provide arguments on significance of this finding 
and on relevance to other experiments.
\end{abstract}

% 74.25.Op Mixed states, critical fields, and surface sheaths  
% 74.25.Bt Thermodynamic properties  
% 74.25.Jb Electronic structure  
%
\maketitle

%\section{Introduction and back ground}
Much attention has been focused on exotic superconductors, 
ranging from high $T_{\rm c}$ cuprates, Ce and U based heavy Fermion materials, filled skutterdites
such as PrOs$_{4}$Sb$_{12}$ to cobaltites 
Na$_{x}$CoO$_{2}$$\cdot$yH$_{2}$O\cite{thal,sigrist}.
The identification of the Cooper-pair symmetry consists 
of two parts; its parity and the gap structure.
The former is responsible for the spin structure of a pair which is
either singlet or triplet. This can be probed by directly measuring the spin susceptibility 
through NMR Knight shift experiment under an applied field.
The latter gap structure is related to the orbital symmetry of a Cooper pair.
This can be probed by thermodynamic measurements 
via a variety of experimental methods, such as temperature ($T$) dependence
of specific heat $C(T)$,
thermal conductivity $\kappa (T)$
or nuclear relaxation time $T_1 (T)$ in NMR-NQR experiments.

The basic principle of this identification for the gap structure is
based on the fact that the energy ($E$) dependence 
of the density of states (DOS) $N(E)$ 
near the Fermi level, which characterizes low-lying excitations of 
a given gap structure.
This $N(E)$ gives rise to a specific power law temperature
dependence\cite{sigrist}. For example, the line (point) node gap yields a $C(T)/T\sim T$($T^2$)
behavior in specific heat, $\kappa (T)\sim T^2 $ ($T^3$) for thermal conductivity
and $T_1^{-1} \sim T^3$ ($T^5$) in nuclear relaxation time at lower $T$ region.
This comes from the fact that the density of states is described by a 
specific functional form; $N(E)\propto |E|$ for line node and $N(E)\propto E^2$ for point node,
through which a simple power counting rule yields specific power law indices 
in various quantities.  Therefore it is decisive to
precisely understand the DOS form $N(E)$ in order to identify the gap structure.
We have attained a lot of information of the pairing symmetry
in various superconductors by this way\cite{sigrist}.

We have been realizing, however, 
that there is no detailed study to investigate 
``the general rule''
for describing $N(E)$ in the mixed state 
of type II superconductor
for various gap structures\cite{fetter}.
Namely although the above thermodynamic measurements are often
performed under an applied field, which is particularly true for $T_1$ in
NMR, one needs to compute the precise functional form 
of $N(E)$ in the vortex state for various gap structures.

In the previous theoretical-work
we studied only the full gap case and analyzed $C(T)$ at low $T$\cite{nakai1}.
However,
this behavior should be examined with including the contribution
of the gap anisotropy.
Usually, experimental data
taken under a field are often analyzed by simply extending the zero field
results discussed above to a finite field $(H>0)$,
keeping the same power law of $N(E)$ with a finite residual DOS $N_0(H)$.
Namely, it has been implicitly
postulated that \mbox{$N(E)=N_0(H)+\alpha |E|^{\gamma}$} 
with $\gamma=0$ (U-shape) in the full gap, $\gamma=1$ in the line node
and $\gamma=2$ in the point node cases.
However, these forms of DOS spectrum should be exactly evaluated
by microscopic calculation without ambiguous assumption.
The functional form of the DOS is important
when we discuss the physical quantities in the mixed state.

Here we calculate the density of states $N(E)$ averaged over a unit cell of a 
vortex lattice in type II superconductors with full gap, 
point and line node gap structures. 
Our microscopic computation takes into account 
exactly
quasi-particle contributions due to vortices, which is not captured by
Ginzburg-Landau and London theories. The results are of the general form
\mbox{$N(E)=N_0(H)+\alpha|E|+O(E^2)$}
for all cases, i.e. $\gamma=1$,
showing that this singular V-shape DOS is universal 
independent of the underlying original gap structure. 

We measured the DOS $N(E)$ for two superconductors;
NbSe$_2$\cite{hess} and YNi$_2$B$_2$C\cite{nishida}
by performing scanning tunneling
spectroscopy (STS) at low temperatures.
We integrate the measured local DOS (LDOS) 
over a certain area around a vortex core.
The former material is known to posses an anisotropic gap
without node while the latter is speculated to be a point-like nodal gap.
Thus these materials provide an excellent testing ground to 
check the theoretical prediction. Indeed both give rise to a V-shape
DOS in common.
This study may be useful in understanding vortices recently 
produced in ultra-cold Fermionic superfluids in $^6$Li atomic gases\cite{mit}.

%\section{The quasi-classical Eilenberger theory}
%
%
{\it Theory}:
The quasiclassical theory is valid in the case of \mbox{$k_{\rm F}\xi\gg 1$},
which is satisfied in almost all type II superconductors.
$k_{\rm F}$ is Fermi-wave number 
and $\xi$ is BCS coherence length which is our units of the length scale.
We introduce the pair potential
\mbox{$\Delta({\bf r})$}, 
the vector potential 
\mbox{${\bf A}({\bf r})$}
and the quasiclassical Green's functions
\mbox{$g({\rm i}\omega_n, {\bf r}, \hat{\bf k})$}, 
\mbox{$f({\rm i}\omega_n, {\bf r}, \hat{\bf k})$}
and \mbox{$f^{\dagger}({\rm i}\omega_n, {\bf r}, \hat{\bf k})$}, 
where ${\bf r}$ is center of mass coordinate of the Cooper pair
and the direction of the momentum 
\mbox{$\hat{\bf k}={\bf k}/\left|{\bf k}\right|$}.
The Eilenberger equation is given by
\begin{eqnarray}
\left\{\omega_n+\frac{\rm i}{2}{\bf v}_{\rm F}\cdot\left(\frac{\nabla}{\rm i}
+\frac{2\pi}{\phi_0}{\bf A}({\bf r})
\right)\right\}f
&=&\Delta({\bf r}, \hat{\bf k})g,
\nonumber 
\\ 
\left\{\omega_n-\frac{\rm i}{2}{\bf v}_{\rm F}\cdot\left(\frac{\nabla}{\rm i}
-\frac{2\pi}{\phi_0}{\bf A}({\bf r})
\right)\right\}f^{\dagger}
&=&\Delta^{\ast}({\bf r}, \hat{\bf k})g,
\label{eq1}
\end{eqnarray}
where
$g=[1-f^{\dagger}f]^{1/2}$, ${\rm Re}\, g>0$, 
\mbox{${\bf v}_{\rm F}=v_{\rm F}\hat{\bf k}$}
is Fermi velocity, and 
$\phi_0$ is a flux quantum\cite{ichioka,nakai2}.
The applied field ${\bf H}$ is along ${\bf z}$-direction.
With the symmetric gauge,
the vector potential is written as
\mbox{${\bf A}({\bf r})=
\frac{1}{2}{\bf H}\times{\bf r}+{\bf a}({\bf r})$}, 
and an internal field ${\bf h}({\bf r})$ is given by 
\mbox{${\bf h}({\bf r})=\nabla\times{\bf a}({\bf r})$}. 
By the numerical calculation
Eq. (\ref{eq1}) is self-consistently solved
with the assumption 
$V_0({\hat{\bf k}},{\hat{\bf k}'})=
V_0\varphi({\hat{\bf k}})
\varphi({\hat{\bf k}'})$
and
\mbox{$\Delta({\bf r}, \hat{\bf k})=
\Delta({\bf r})\varphi(\hat{\bf k})$}, 
considering 
the self-consistent conditions for
\mbox{$\Delta({\bf r})$} 
and
\mbox{${\bf a}({\bf r})$}; 
\begin{eqnarray}
\Delta({\bf r})
&=&N_02\pi T\sum^{\omega_c}_{\omega_n>0}
V_0
\langle
\varphi({\hat{\bf k})}
f({\rm i}\omega_n, {\bf r}, \hat{\bf k})
\rangle_{\hat{\bf k}},
\label{scD} 
\\
{\bf j}({\bf r})&=&\frac{\pi\phi_0}{\kappa^2\Delta_0\xi^3}
2\pi T\sum^{\omega_c}_{\omega_n>0}
{\rm i}
\langle
\hat{\bf k}g({\rm i}\omega_n, {\bf r}, \hat{\bf k})
\rangle_{\hat{\bf k}}
,
\label{scA} 
\end{eqnarray}
where 
${\bf j}({\bf r})=\nabla\times\nabla\times {\bf a}({\bf r})$, 
$N_0$ is density of states
at the Fermi level in the normal state.
$\langle\cdots\rangle_{\hat{\bf k}}$ means
the average over the direction of $\hat{\bf k}$.
The cut-off energy is set as \mbox{$\omega_{\rm c}=20T_{\rm c}$}.
$\Delta_0$ is uniform gap at $T=0$.
We use the high $\kappa$ for the Ginzburg-Landau parameter.
The angle dependence of the pair potential $\varphi({\hat{\bf k}})$
specifies various gap structures with isotropic, 
line node or point node gaps.

The LDOS at an energy $E$
is given by
\begin{eqnarray}
N(E,{\bf r})=N_0 
\langle
{\rm Re}\, g({\rm i}\omega_n\to E+{\rm i}\eta,{\bf r}, \hat{\bf k})
\rangle_{\hat{\bf k}},
\label{N} 
\end{eqnarray}
where $g$ is calculated by Eq. (\ref{eq1}) 
with \mbox{${\rm i}\omega_n\to E+{\rm i}\eta$}. 
We set \mbox{$\eta=0.01\Delta_0$}.
The total DOS $N(E)$ is the spatial average of the LDOS, i.e. 
$N(E)= \langle N(E,{\bf r}) \rangle_{\bf r}$. 
The self-consistent calculation is performed 
within the vortex lattice unit cell,
which is divided into \mbox{$81 \times 81$} mesh points.
We assume that vortices form a triangular lattice.
The cylindrical Fermi-surface is chosen 
for the $s$-wave and the line-node case. 
And we use the spherical Fermi-surface for the point-node case.

\begin{figure}[tb]
\includegraphics[keepaspectratio]{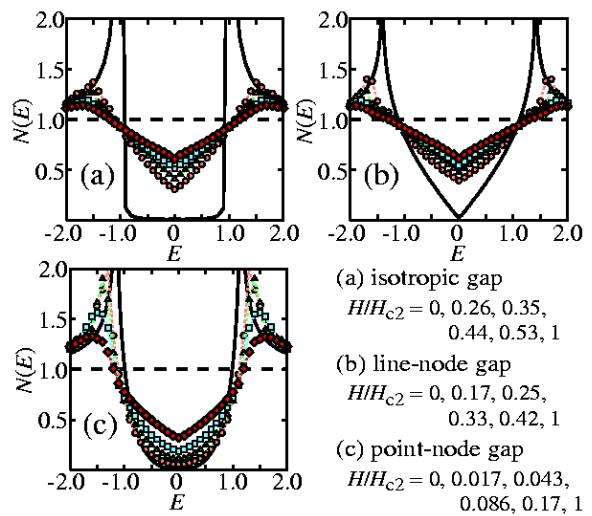}%
\caption{(Color online)
Averaged density of states $N(E)$ of (a) isotropic gap, (b) line-node gap
and (c) point-node gap.
$H$ increases
from bottom to top at $E=0$.
}
%\label{}
\end{figure}
We display the DOS $N(E)$ for the $s$-wave case with the 
isotropic gap in Fig. 1(a).
It is seen from this that the DOS has a V-shape form with
a singularity at $E=0$, namely, 
\begin{eqnarray}
N(E)=N_0(H)+ \alpha {|E|\over \Delta_0}+O(E^2)
\label{} 
\end{eqnarray}
\noindent
where $N_0(H)$ is the energy independent term, corresponding to
the zero-energy DOS which is only a function of the field $H$.
This yields the Sommerfeld $T$-linear coefficient $\gamma_0(H)$
in the specific heat as
$\gamma_0(H) \equiv
\lim_{T\rightarrow 0}C(T)/T=(2\pi^2/3)N_0(H)$.
We confirm that the sharp V-shape DOS is easily smeared 
by thermal broadening effect. 
Impurities also affect smearing of
the sharp edge of the V-shape DOS  which becomes round 
by finite mean free path effect
due to impurities.
The linear slope $\alpha(H)$ in $N(E)$ is field dependent;
For the $s$-wave in Fig. 1(a), 
$\alpha(H)$ is small at lower fields.
As $H$ increases $\alpha(H)$ takes a maximum around $H/H_{\rm c2}\sim 0.3$ 
and then becomes smaller
again towards $H_{\rm c2}$, continuing to a flat DOS in the normal state.
This non-monotonic behavior of the slope is the same also for
the point node case, but is contrasted with that in the
line node case, which are discussed below.

In Fig. 1(b) we exhibit the results for the line node case  described by 
the pair function
\mbox{$\varphi(\theta, \phi)=\sqrt{2}\cos2\theta$}. At $H=0$,
$N(E)=\alpha |E|/\Delta_0$ without a constant term in the low energy
range as expected. In the vortex state this expression is 
valid except that we must add a field dependent $E=0$ value,
that is, $N(E)=N_0(H)+ \alpha {|E|/\Delta_0}$.
At first sight this result seems ``apparent'' because the  original
zero field DOS has a V-shape and it  is simply shifted up-wards.
However, this is not the case because the V-shape DOS is obtained even
when the original gap structure is not a V-shape as already shown in the $s$-wave case.
The field dependence of $\alpha (H)$ in the line node case is seen to be 
a monotonic decreasing function with $H$ where the maximum slope occurs at $H=0$.
In Fig. 1(c) we show the V-shape DOS for the point-node case at $H>0$.
It is seen that at low fields the linear portion of $|E|$ is limited
to low energies around $E=0$. But as $H$ increases a V-shape DOS
feature manifests itself for wider energy ranges,
so that the form in Eq. (5) is evident.

\begin{figure}[tb]
\includegraphics[keepaspectratio]{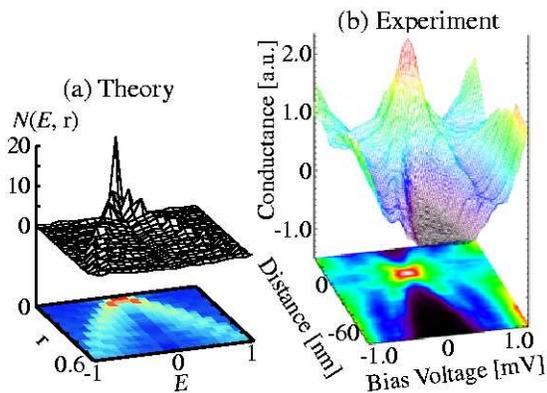}%
\caption{(Color online)
(a) Theoretical calculation of LDOS $N(E, r)$, showing very similar sub gap peak structures to (b).
(b) The spectral evolution $N(E, r)$ in NbSe$_2$ at 0.025 T along 109 nm length that intersects a vortex as measured by STM-STS.  
}
%\label{}
\end{figure}
This ubiquitous V-shape DOS can be understood in terms of the LDOS.
We show the so-called spectral evolution for the $s$-wave case in Fig. 2(a)
where the spectral weight  $N(E, r)$ in Eq. (4) is displayed
in a plane of the distance $r$ from the vortex core and the energy $E/\Delta_0$.
The zero-energy peak at the core site is split into two peaks as moving 
away from the core, that is, the  two ``trajectories'' are given by $r=\beta |E|$.
This spectral evolution agrees with the STS observation on NbSe$_2$
shown in Fig. 2(b).
The total DOS is obtained by integrating this spectral weight 
spatially for each energy, namely,
\begin{eqnarray}
N(E)=N_0+\int^\infty_0dr r \delta (r-\beta |E|)=N_0+\beta^2|E|.
\label{} 
\end{eqnarray}
\noindent
The first term comes from the zero-energy peak at $r=0$.
This argument is applicable for the $d$-wave case
where instead of the circular symmetry for the $s$-wave
case the four fold symmetry must be taken account, which 
amounts to modifying the coefficient $\beta^2$,
but the $|E|$ dependence remains unchanged.
Thus it is understood that the V-shape DOS is universal, independent of the
underlying original gap structure.

{\it Experiment}:
NbSe$_2$ is a typical anisotropic $s$-wave superconductor with a gap ranging from 0.7 to 1.3 meV, $T_{\rm c}$ = 7.2 K and $H_{\rm c2}(T=0)$ = 3.2 T, whose layered structure and Van der Waals surfaces are ideal for STM-STS experiments\cite{hess}.  When an external field is applied to the $c$-axis, the vortices form a triangular lattice.  A series of differential conductance spectra $\sigma(r,V)$ are taken along a line that extends radially through a vortex. Such an example taken at 50 mK and 0.025 T is shown in Fig. 2(b). The overall features and numerous details of the LDOS agree very well with the theoretical calculations  shown in Fig. 2(a) (also see Ref. 10).

\begin{figure}[tb]
\includegraphics[keepaspectratio]{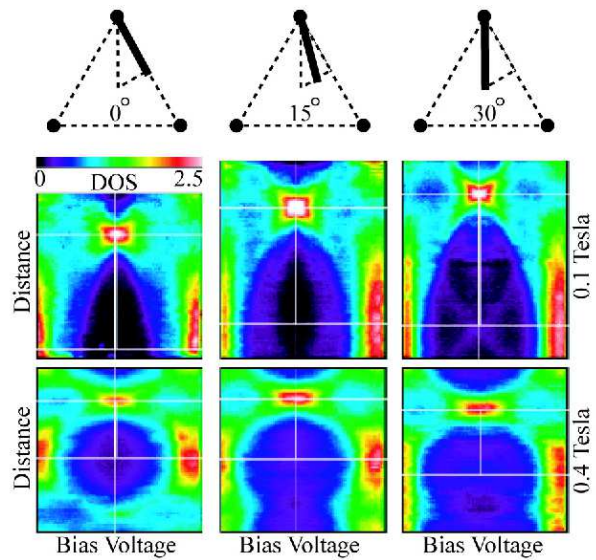}%
\caption{(Color online)
Spectral evolution along 3 paths (black solid lines) for NbSe$_2$ with respect to the vortex lattice for $H=0.1$ T and 0.4 T. The LDOS data between the 2 horizontal white lines are used in the integration.
}
%\label{}
\end{figure}
In order to deduce the spatially averaged DOS over a unit cell, the data are collected along the line paths that extend out in three different directions from a vortex as shown in Fig. 3.  The normalized DOS are inferred by the ratio of conductance, 
$\sigma(r,V)/\sigma(r,V=5\,\,{\rm meV})$, and are shown as a color scale image. This data is weighted by radius and summed to emulate an integration over the unit cell.   This is essentially a polar coordinate integration: 
$N(E) = \int N(E, r,\theta)r dr d\theta$.  The  resulting averaged DOS are displayed in Fig. 4(a) for two fields $H$= 0.1 T and  0.4 T along 
with the DOS at a zero field showing the $s$-wave gap function. 
A definite V-shape in the DOS at finite fields is evident
between $-0.6$ and $0.6$ meV, and confirmed by the better fit attained 
by $\sigma(V)=\sigma_0+\alpha |V|+\beta V^2$ (solid line) 
over that of a simpler parabolic form $\sigma(V)=\sigma_0+\beta V^2$
(dashed line) as shown in Fig. 4(b).  Furthermore we note that the zero bias offset $\sigma/\sigma_{\rm N}$ ($\sigma_{\rm N}$ is the normal state conductance) corresponding to the Sommerfeld coefficient $\gamma(H)/\gamma_{\rm N}$ at low $T$ nicely matches the specific heat experiments\cite{hanaguri} to within 5$\%$. 

\begin{figure*}[tb]
\includegraphics[keepaspectratio]{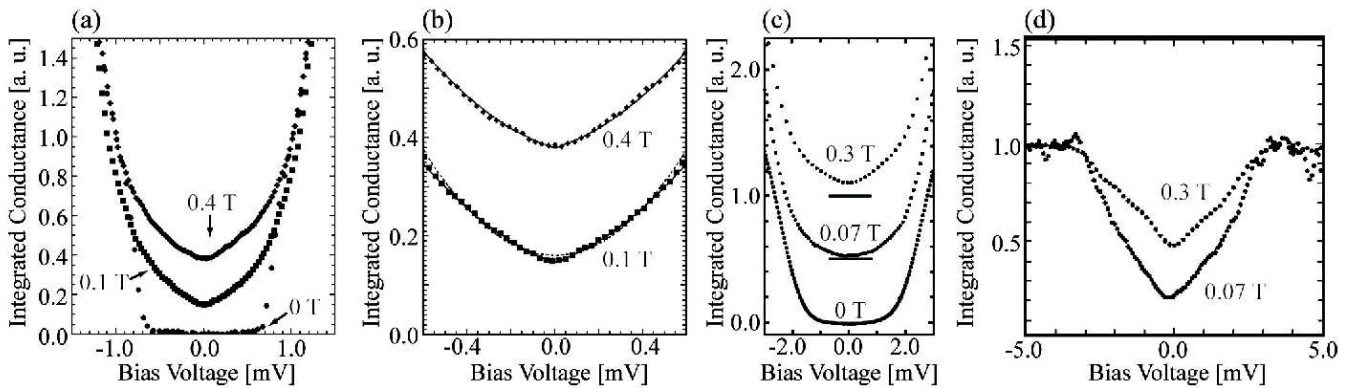}%
\caption{
(a) Spatially averaged DOS $N(E)$ over a unit cell at $H$=0, 0.1 T and 0.4 T
for NbSe$_2$.
(b) Same data but over $\pm$0.6 mV range inside the gap.
The solid lines with $a+b|V|+c|V|^2$ are seen better fitting curves
than the dash line with a quadratic form $a+c'|V|^2$.
(c) DOS $N(E)$ averaged over a unit cell at $H$=0, 0.07 T and 0.3 T
for YNi$_2$B$_2$C.
(d) DOS $N(E)$ averaged over circular area with diameter 10 nm around a core
in YNi$_2$B$_2$C.
}
%\label{}
\end{figure*}
Similar analyses have been performed on YNi$_2$B$_2$C whose gap structure is speculated 
to be point-like nodes and $\sigma(r,V)$ has been measured at 460 mK\cite{nishida}.
The samples have $T_{\rm c}$= 15.6 K and $H_{\rm c2}$($T=0$)= 8 T.
When external field is applied to the $c$-axis of this tetragonal system,
a square vortex lattice is formed.
At $T=450$ mK, STS data are collected for various spatial points and 
bias voltages to yield the average DOS $N(E)$.
The results are shown in Figs. 4(c) and (d) for $H=0$, 0.07 T and 0.3 T.
The conductance data sets for finite fields are seen to
posses a V-shape dependence. Although it is difficult from these data
to determine the precise functional form $\sigma(V)=\sigma_0+\alpha |V|^{\eta}$,
it contains a linear term $\eta=1$.

As is clear from the above argument leading to Eq. (6), we could see a V-shape tunneling
conductance if we restrict our integration region to a narrower region around a core.
This procedure emphasizes the contribution from the
vortex core which ultimately yields a  V-shape DOS. In fact
as shown in Fig. 4(d) which is obtained by integrating
over a circular area with diameter 10 nm centered at a core, 
one can see a clear V-shape conductance curve for both fields.

The ubiquitous V-shape DOS has important consequences 
on identifying the pairing symmetry, in particular,  
its gap structure through the $T$-dependences of various thermodynamic 
quantities as mentioned before.
At low temperatures the $T$ dependence is governed 
by the functional form of DOS.
Using the V-shape DOS in Figs. 1 and 4,
the simple power law counting  tells us the followings:
specific heat $C(T)/T=\gamma_0(H)+\alpha' T$,
nuclear relaxation time $T_1(T)^{-1}\propto T+c'T^3$,
and thermal conductivity $\kappa(T)\propto T+c''T^2$.
Therefore, in the experiment under magnetic field, 
we can not simply assign the origin of the power law behavior 
as a line node,   
since the origin may be the V-shape DOS due to the vortex states,
not due to the line node gap structure. 
The $T_1$-behavior $T_1(T)^{-1}\propto T^3$ under magnetic fields 
is a result of the spatial average. 
If we observe $T_1$ outside of the vortex core 
by site selective NMR technique\cite{takigawa}, 
we can unambiguously identify the signal due to the line node.

Simon and Lee\cite{simon}, and Won and Maki\cite{won} derive a scaling law to describe
$T$- and $H$-dependences of various thermodynamic quantities
for the line node $d$-wave systems. In particular,
the latter authors explicitly find a scaling function 
based on the averaged DOS obtained from the Doppler shift idea\cite{volovik}
whose functional form differs from ours.
The Simon-Lee and Won-Maki scaling, which is quite successful for wide range of line node superconductors
(see for example, Ref. 16),
can be improved and applied to other superconductors 
with different gap structures
by using the obtained V-shape DOS.

In summary, 
by the self-consistent quasiclassical calculation
evaluating the vortex contribution exactly,
we have demonstrated the averaged density of states 
in the mixed state is  a V-shape, described by
\mbox{$N(E)=N(E=0)+\alpha_E{|E|/\Delta_0}+O(E^2)$}. 
This formula is valid for any underlying gap structures; 
isotropic, point node or line node gaps.
However the $\alpha_E$ is small 
for the isotropic and point-node gap case in a low field.
Two STM experiments on NbSe$_2$ and YNi$_2$B$_2$C unambiguously
exhibit this behavior in their tunneling conductance.
The vortex-lattice geometry does not affect the V-shape 
and just slightly changes $\alpha_E$.
We have discussed several important consequences.
In particular, when one identifies the gap symmetry by
measuring thermodynamic quantities, careful
consideration is needed in the mixed state.
%
%
%\begin{acknowledgments}
%\end{acknowledgments}

\end{document}